# Manipulating type-I and type-II Dirac polaritons in cavity-embedded honeycomb metasurfaces

Charlie-Ray Mann  [1], Thomas J. Sturges  [1], Guillaume Weick[2], William L. Barnes  [1] & Eros Mariani[1]

Pseudorelativistic Dirac quasiparticles have emerged in a plethora of artificial graphene systems that mimic the underlying honeycomb symmetry of graphene. However, it is notoriously difficult to manipulate their properties without modifying the lattice structure. Here we theoretically investigate polaritons supported by honeycomb metasurfaces and, despite the trivial nature of the resonant elements, we unveil rich Dirac physics stemming from a non-trivial winding in the light–matter interaction. The metasurfaces simultaneously exhibit two distinct species of massless Dirac polaritons, namely type-I and type-II. By modifying only the photonic environment via an enclosing cavity, one can manipulate the location of the type-II Dirac points, leading to qualitatively different polariton phases. This enables one to alter the fundamental properties of the emergent Dirac polaritons while preserving the lattice structure—a unique scenario which has no analog in real or artificial graphene systems. Exploiting the photonic environment will thus give rise to unexplored Dirac physics at the subwavelength scale.

[1] EPSRC Centre for Doctoral Training in Metamaterials (XM2), Department of Physics and Astronomy, University of Exeter, Exeter EX4 4QL, UK. [2] Université de Strasbourg, CNRS, Institut de Physique et Chimie des Matériaux de Strasbourg, UMR 7504, F-67000 Strasbourg, France. Correspondence and requests for materials should be addressed to C.-R.M. (email: cm433@exeter.ac.uk) or to E.M. (email: E.Mariani@exeter.ac.uk)

The groundbreaking discovery of monolayer graphene[1] has inspired an extensive quest to emulate massless Dirac quasiparticles in a myriad of distinct artificial graphene systems[2–11], ranging from ultracold atoms in optical lattices[3] to evanescently coupled photonic waveguide arrays[4]. Owing to their honeycomb symmetry, linear band-degeneracies manifest in the quasiparticle spectrum which we call conventional Dirac points (CDPs). These belong to the ubiquitous type-I class of two-dimensional (2D) Dirac points that are characterized by Dirac cones with closed isofrequency contours. As a result, the corresponding quasiparticles are described by the rather exotic 2D massless Dirac Hamiltonian[12], and thus offer fundamental insight into pseudorelativistic phenomena such as the iconic Klein paradox[13]. The latter is responsible for the suppression of backscattering and for the antilocalization of massless Dirac quasiparticles, which are highly desirable properties for efficient quasiparticle propagation in novel devices.

Since the existence of type-I CDPs is intrinsically linked to the honeycomb structure, the fundamental properties of the massless Dirac quasiparticles are notoriously robust and difficult to manipulate. However, by exploiting meticulous control over the lattice structure, artificial graphene systems have enabled the exploration of Dirac quasiparticles in new regimes that are difficult, if not impossible to achieve in graphene itself[14–19]. Among others, the archetypal example which has attracted considerable interest is the paradigm of strain-engineering, where it has been shown that lattice anisotropy can induce the merging and annihilation of type-I CDPs[3,14–16,20–23], and that aperiodicity can generate large pseudomagnetic fields[17,24].

Moreover, the recent discovery of type-II Dirac/Weyl semimetals[25–29] sparked a burgeoning exploration into the prospects of a rarer type-II class of three-dimensional Dirac/Weyl points. As the latter are characterized by critically tilted Dirac/Weyl cones with open, hyperbolic isofrequency contours, the corresponding Lorentz-violating Dirac/Weyl quasiparticles exhibit markedly different properties from their type-I counterparts[25–29]. Soon after their realization, electromagnetic analogs emerged[30–34], and this exploration has recently been extended to 2D systems where a distinct type-II class of 2D Dirac points were theoretically predicted[35,36]. However, since their existence is predicated on strong anisotropy in judiciously engineered photonic structures, one cannot manipulate their properties without modifying the lattice structure.

This hunt for exotic quasiparticles has recently entered the realm of polaritonics[37–42]. The true potential of polaritons lies in their hybrid nature, where their light and matter constituents can be manipulated independently, thereby providing additional tunable degrees of freedom. Among other examples, recent works have shown the tantalizing prospect of engineering novel topological polaritons by introducing a winding coupling between ordinary photons and excitons[39,41].

In this work, we exploit the hybrid nature of polaritons in a different setting, namely metasurfaces, and we unveil unique Dirac physics by shifting the focus from the lattice structure and its deformations to the effect of manipulating the surrounding photonic environment. In particular, we theoretically study the polaritons supported by imminently realizable, crystalline metasurfaces consisting of a honeycomb array of resonant, dipolar meta-atoms. Despite the elementary nature of these metasurfaces, we unveil the simultaneous existence of both type-I and type-II massless Dirac polaritons which have distinct physical origins. Crucially, the existence of the latter is not a result of anisotropy but is intrinsically linked to the hybrid nature of the polaritons, emerging from a non-trivial winding in the light–matter interaction. Furthermore, we show that by embedding the honeycomb metasurface inside a planar photonic cavity and simply changing the cavity height, one can induce multiple phase transitions including the multimerging of type-I and type-II Dirac points and the annihilation of type-II Dirac points. This striking tunability results in qualitatively different polariton phases, despite the preserved lattice structure. In particular, we unveil a morphing between a linear and a parabolic spectrum accompanied by a change in the topological Berry phase, and an environment-induced inversion of chirality, all of which have no analog in graphene or artificial graphene systems studied thus far. Therefore, this unique paradigm will give rise to unexplored Dirac-related phenomena at the subwavelength scale, such as anomalous Klein tunneling, negative refraction, and pseudomagnetic Landau levels, which can all be tuned via the photonic environment alone.

## Results

**Hamiltonian formulation**. While metamaterials have traditionally been described in terms of macroscopic effective properties[30,33,43], the importance of crystallinity is becoming increasingly apparent[44]. Therefore, to capture the essential physics related to complex non-local effects that arise from strong multiple-scattering[45], here we study the properties of the cavity-embedded honeycomb metasurface by means of a microscopic Hamiltonian formalism. This allows us to clearly identify the distinct physical origins of the type-I and type-II Dirac points.

The full polariton Hamiltonian of this system reads $H_{pol} = H_{mat} + H_{ph} + H_{int}$, where the interaction Hamiltonian $H_{int}$ couples the matter and photonic subspaces whose free dynamics are governed by $H_{mat}$ and $H_{ph}$, respectively. We employ the Coulomb gauge, where the instantaneous Coulomb interaction between the meta-atoms is incorporated within the matter Hamiltonian $H_{mat}$, and the effects of the dynamic photonic environment—described by the transverse vector potential—are included through the principle of minimal-coupling[46].

A schematic of a cavity-embedded honeycomb metasurface is depicted in Fig. 1. We model each subwavelength meta-atom by a single dynamical degree of freedom describing the electric-dipole moment associated with its (non-degenerate) fundamental eigenmode with resonant frequency $\omega_0$. These meta-atoms are then oriented such that their dipole moments point normal to the plane of the lattice. Furthermore, we consider subwavelength nearest-neighbor separation $a$ such that the light cone intersects the Brillouin zone edge above $\omega_0$, ensuring the existence of evanescent bound, subwavelength polaritons. The strength of the Coulomb dipole–dipole interaction between neighboring meta-atoms is parametrized by $\Omega$. Finally, the metasurface is embedded at the center of a planar photonic cavity of height $L$, where the cavity walls are assumed to be lossless and perfectly conducting metallic plates. Such a structure is imminently realizable across the electromagnetic spectrum from arrays of plasmonic nanoparticles to microwave helical resonators (see Fig. 1).

**Emergence of type-I Dirac points**. The matter Hamiltonian within the nearest-neighbor approximation reads

$$H_{mat} = \hbar\tilde{\omega}_0 \sum_{\mathbf{q}} \left( a_{\mathbf{q}}^{\dagger} a_{\mathbf{q}} + b_{\mathbf{q}}^{\dagger} b_{\mathbf{q}} \right) + \hbar\tilde{\Omega} \sum_{\mathbf{q}} \left( f_{\mathbf{q}} b_{\mathbf{q}}^{\dagger} a_{\mathbf{q}} + \text{H.c.} \right), \quad (1)$$

where, for brevity, we have not presented the non-resonant terms (see Methods for derivation). In Eq. (1), $\tilde{\omega}_0$ is the renormalized resonant frequency and $\tilde{\Omega}$ is the renormalized Coulombic interaction strength due to the cavity-induced image dipoles (see Methods for their dependence on the cavity height). The bosonic operators $a_{\mathbf{q}}^{\dagger}$ and $b_{\mathbf{q}}^{\dagger}$ create quanta of the quasistatic

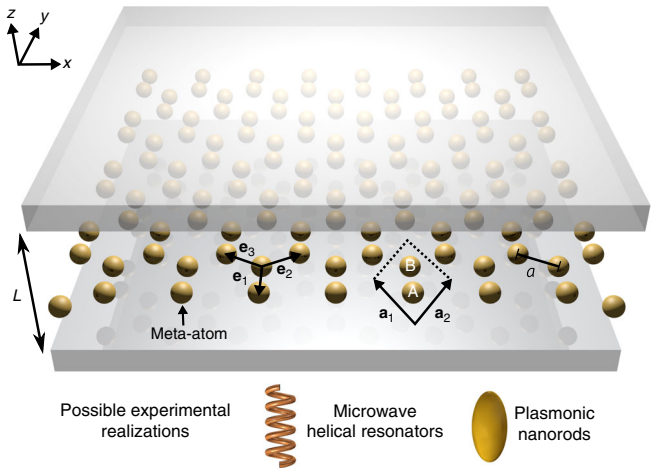

**Fig. 1** Schematic of a cavity-embedded honeycomb metasurface. The honeycomb array of meta-atoms is composed of two inequivalent (A and B) hexagonal sublattices—defined by lattice vectors $\mathbf{a}_1 = a(-\frac{\sqrt{3}}{2}, \frac{3}{2})$ and $\mathbf{a}_2 = a(\frac{\sqrt{3}}{2}, \frac{3}{2})$—which are connected by nearest-neighbor vectors $\mathbf{e}_1 = a(0, -1)$, $\mathbf{e}_2 = a(\frac{\sqrt{3}}{2}, \frac{1}{2})$, and $\mathbf{e}_3 = a(-\frac{\sqrt{3}}{2}, \frac{1}{2})$, where $a$ is the subwavelength nearest-neighbor separation. Each subwavelength meta-atom is modeled as an electric dipole, oriented normal to the plane of the lattice. The honeycomb metasurface is then embedded inside a photonic cavity of height $L$, which is composed of two perfectly conducting metallic plates, enabling one to modify the photonic environment while preserving the lattice structure. This general model can be readily realized across the electromagnetic spectrum, from arrays of plasmonic nanorods to microwave helical resonators

collective-dipole modes that extend across the A and B sublattices, respectively, with wavevector $\mathbf{q}$ in the first Brillouin zone (see Fig. 2a). Finally, the function $f_{\mathbf{q}} = \sum_{j=1}^{3} \exp(i\mathbf{q} \cdot \mathbf{e}_j)$ encodes the honeycomb geometry of the lattice with nearest-neighbor vectors $\mathbf{e}_j$ (see Fig. 1).

We diagonalize the matter Hamiltonian (Eq. (1)) as $H_{mat} = \sum_{\tau = \pm} \sum_{\mathbf{q}} \hbar\omega_{\mathbf{q}\tau}^{mat} \beta_{\mathbf{q}\tau}^{\dagger} \beta_{\mathbf{q}\tau}$ where the bosonic operators $\beta_{\mathbf{q}\tau}^{\dagger} = \psi_{\mathbf{q}}^{\dagger} |\psi_{\mathbf{q}\tau}\rangle$ create quasistatic collective-dipole normal modes with dispersion $\omega_{\mathbf{q}\tau}^{mat} = \tilde\omega_0 + \tau\tilde\Omega |f_{\mathbf{q}}|$. Here, $\tau$ indexes the upper ($\tau = +1$) and lower ($\tau = -1$) bands and $\psi_{\mathbf{q}}^{\dagger} = (a_{\mathbf{q}}^{\dagger}, b_{\mathbf{q}}^{\dagger})$ is a spinor creation operator. The spinors $|\psi_{\mathbf{q}\tau}\rangle = (1, \tau e^{i\varphi_{\mathbf{q}}})^T/\sqrt{2}$, where $^T$ denotes the transpose, describe an emergent pseudospin degree of freedom where the two components encode the relative amplitude and phase of the dipolar oscillations on the two inequivalent A and B sublattices, respectively, with $\varphi_{\mathbf{q}} = \arg(f_{\mathbf{q}})$. These spinors can be represented by a pseudospin vector on the Bloch sphere which reads $\mathbf{S}_{\mathbf{q}\tau} = \tau(\cos\varphi_{\mathbf{q}}, \sin\varphi_{\mathbf{q}}, 0)$.

At the high symmetry K and K' points (see Fig. 2a), the sublattices decouple with no well-defined relative phase (i.e., $f_{\mathbf{q}} = 0$), giving rise to two inequivalent CDPs located at $\pm \mathbf{K} = \pm (\frac{4\pi}{3\sqrt{3}a}, 0)$ as observed in Fig. 2b. These CDPs correspond to vortices in the pseudospin vector field $\mathbf{S}_{\mathbf{q}\tau}$, which give rise to topological singularities in the Berry curvature[47]. Therefore, the CDPs are sources of quantized Berry $w\pi$, where $w = \pm 1$ is the topological charge of the Dirac point corresponding to the winding number of the vortex. As expected from the symmetry of the metasurface, the existence of the CDPs is robust against long-range Coulomb interactions as shown in Supplementary Note 1. In fact, for small cavity heights, the image dipoles quench long-range Coulomb interactions and the nearest-neighbor approximation becomes increasingly accurate as shown in Supplementary Figure 1.

To quadratic order in $\mathbf{k} = \mathbf{q} - \mathbf{K}$ ($ka \ll 1$), the effective matter Hamiltonian near the K point is $H_{\mathbf{K}}^{eff} = \sum_{\mathbf{k}} \psi_{\mathbf{k}}^{\dagger} \mathcal{H}_{\mathbf{K},\mathbf{k}}^{eff} \psi_{\mathbf{k}}$, with spinor creation operator $\psi_{\mathbf{k}}^{\dagger} = (a_{\mathbf{k}}^{\dagger}, b_{\mathbf{k}}^{\dagger})$ and Bloch Hamiltonian

$$\mathcal{H}_{\mathbf{K},\mathbf{k}}^{eff} = \hbar\tilde\omega_0 \mathbb{1}_2 - \hbar\tilde{v}\boldsymbol{\sigma} \cdot \mathbf{k} + \hbar\tilde{t}(\boldsymbol{\sigma}^* \cdot \mathbf{k})^{\circ 2} \qquad (2)$$

Here, $\mathbb{1}_2$ is the $2 \times 2$ identity matrix, $\boldsymbol{\sigma} = (\sigma_x, \sigma_y)$ and $\boldsymbol{\sigma}^* = (\sigma_x, -\sigma_y)$ are vectors of Pauli matrices, and $^{\circ 2}$ represents the Hadamard (element-wise) square. Note that the image dipoles do not qualitatively affect the physics, but simply lead to a renormalization of the group velocity $\tilde{v} = 3\tilde\Omega a/2$ and trigonal warping parameter $\tilde{t} = 3\tilde\Omega a^2/8$. Apart from a global energy shift, Eq. (2) is equivalent to a 2D massless Dirac Hamiltonian to leading order in $\mathbf{k}$, with an isotropic Dirac cone spectrum $\omega_{\mathbf{k}\tau}^{mat} = \tilde\omega_0 + \tau\tilde{v}|\mathbf{k}|$ that is characterized by closed isofrequency contours. Therefore, as expected from the honeycomb symmetry, the CDP belongs to the type-I class of 2D Dirac points, and the corresponding spinors $|\psi_{\mathbf{k}\tau}\rangle = (1, -\tau e^{i\theta_{\mathbf{k}}})^T/\sqrt{2}$, where $\theta_{\mathbf{k}} = \arctan(k_y/k_x)$, represent massless Dirac collective-dipoles with a topological Berry phase of $\pi$. The effective Hamiltonian near the K' point is given by $\mathcal{H}_{\mathbf{K}',\mathbf{k}}^{eff} = (\mathcal{H}_{\mathbf{K},-\mathbf{k}}^{eff})^*$, where the corresponding massless Dirac collective-dipoles have a topological Berry phase of $-\pi$, as required by time-reversal symmetry.

**Hybridization with the photonic environment**. Given the subwavelength nearest-neighbor separation, it is tempting to assert that the near-field Coulomb interactions in $H_{mat}$ capture the essential physics. In fact, we will show that this quasistatic description misses the profound influence of the surrounding photonic environment, which has a remarkably non-trivial effect on the Berry curvature and, therefore, on the corresponding nature of the polaritons.

Crucially, the metallic cavity supports a fundamental transverse electromagnetic (TEM) mode whose polarization (parallel to the dipole moments) and linear dispersion (see Fig. 2b) are independent of the cavity height. For brevity, in what follows we do not present the contributions from the other cavity modes since the essential physics emerges from the interaction with the fundamental TEM mode (see Methods for the full expressions). In fact, the higher order cavity modes become increasingly negligible for smaller cavities as they are progressively detuned from the dipole resonances.

The effects of the photonic environment are encoded in the free photonic Hamiltonian

$$H_{ph} = \hbar \sum_{\mathbf{q}n} \omega_{\mathbf{q}n}^{ph} c_{\mathbf{q}n}^{\dagger} c_{\mathbf{q}n} \qquad (3)$$

and in the light–matter interaction Hamiltonian $H_{int} = H_{int}^{(1)} + H_{int}^{(2)}$, with

$$\begin{aligned} H_{int}^{(1)} = &\hbar \sum_{\mathbf{q}n} i\xi_{\mathbf{q}n} \phi_n^* \left( a_{\mathbf{q}}^{\dagger} c_{\mathbf{q}n} + a_{\mathbf{q}}^{\dagger} c_{-\mathbf{q}n}^{\dagger} \right) \\ &+ \hbar \sum_{\mathbf{q}n} i\xi_{\mathbf{q}n} \phi_n \left( b_{\mathbf{q}}^{\dagger} c_{\mathbf{q}n} + b_{\mathbf{q}}^{\dagger} c_{-\mathbf{q}n}^{\dagger} \right) + \text{H.c.} \end{aligned} \qquad (4)$$

and

$$\begin{aligned} H_{int}^{(2)} = &\hbar \sum_{\mathbf{q}nn'} \frac{2\xi_{\mathbf{q}n}\xi_{\mathbf{q}n'}}{\omega_0} \text{Re}(\phi_n \phi_{n'}^*) \\ &\left( c_{\mathbf{q}n}^{\dagger} c_{\mathbf{q}n'} + c_{\mathbf{q}n}^{\dagger} c_{-\mathbf{q}n'}^{\dagger} \right) + \text{H.c.} \end{aligned} \qquad (5)$$

where $\xi_{\mathbf{q}n} \propto L^{-1/2}$ parametrizes the strength of the light–matter interaction (see Methods for analytical expression). The bosonic operator $c_{\mathbf{q}n}^{\dagger}$ creates a TEM photon with wavevector $\mathbf{q}$ in the first Brillouin zone and dispersion $\omega_{\mathbf{q}n}^{ph} = c|\mathbf{q} - \mathbf{G}_n|$, where $n$ indexes the set of reciprocal lattice vectors $\mathbf{G}_n$. The complex phase factors $\phi_n = \exp(ia\mathbf{G}_n \cdot \hat{\mathbf{y}})$ are associated with Umklapp processes that arise due to the discrete, in-plane translational symmetry of the metasurface, and must be retained as they are critical for maintaining the point-group symmetry of the polariton Hamiltonian.

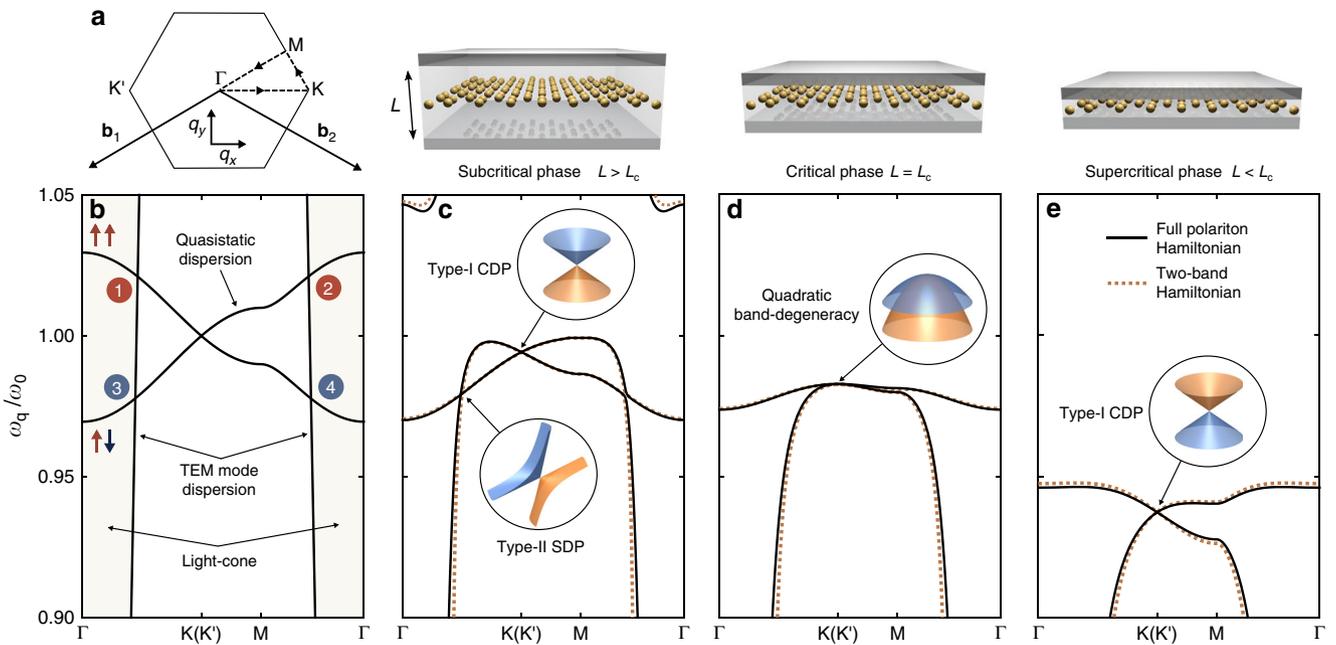

**Fig. 2** Evolution of the polariton dispersion as the cavity height is reduced. **a** First Brillouin zone defined by primitive reciprocal lattice vectors $\mathbf{b}_1 = \frac{2\pi}{3a}(-\sqrt{3}, -1)$ and $\mathbf{b}_2 = \frac{2\pi}{3a}(\sqrt{3}, -1)$. **b** Quasistatic dispersion of the collective-dipole normal modes, where the upper band corresponds to a bright, symmetric dipole configuration (↑↑) and the lower band corresponds to a dark, antisymmetric dipole configuration (↑↓). The light-cone (shaded region) is bounded by the linear dispersion of the TEM mode. Due to the non-trivial winding in the light-matter interaction (see Fig. 3), the band crossings are expected to result in large (band crossings '1' and '2') or small (band crossings '3' and '4') direction-dependent anticrossings in the polariton spectrum. **c–e** Polariton dispersion obtained from the polariton Hamiltonian $H_{pol}$ (solid black lines) and the two-band Hamiltonian $\overline{H}_{mat}$ (orange dashed lines), for **c** subcritical ($L = 5a$), **d** critical ($L = L_c = 1.75a$), and **e** supercritical ($L = a$) cavity heights, respectively. While type-I CDPs with an isotropic Dirac cone (see inset of **c**) exist even in the quasistatic dispersion (see **b**), new type-II SDPs with a critically tilted Dirac cone (see inset in **c**) emerge due to the vanishing light-matter interaction for the dark quasistatic band along the Γ–K(K') directions (see Fig. 3). At the critical cavity height $L_c$, three type-II SDPs merge with the type-I CDP (see Fig. 5) resulting in a quadratic band-degeneracy at K(K') (see inset in **d**). After criticality, the type-II SDPs annihilate one another and the massless Dirac cone re-emerges at the type-I CDPs (see inset in **e**) accompanied by an inversion of chirality (see Fig. 5). Plots obtained with parameters $\omega_{Ko}^{ph} = 2.5\omega_0$ and $\Omega = 0.01\omega_0$

We diagonalize $H_{pol}$ using a generalized Hopfield–Bogoliubov transformation[48] (see Methods for details), and in Fig. 2c–e, we present the resulting polariton dispersion for different cavity heights. Also, in Supplementary Figure 2, we present the full polariton dispersion which includes long-range Coulomb interactions. For small cavity heights, the full polariton dispersion is almost indistinguishable from that obtained in the nearest-neighbor approximation, and therefore one can conclude that long-range Coulomb interactions do not qualitatively affect the physics presented here. It is important to stress that our general model captures the essential physics that will emerge in a variety of different experimental setups. To show this, in Supplementary Figure 3 and Supplementary Figure 4 we present the polariton dispersions obtained from finite element simulations of a honeycomb array of plasmonic nanorods and microwave helical resonators, respectively. Indeed, these entirely different physical realizations show the same evolution of the polariton spectrum as presented in Fig. 2c–e and Supplementary Figure 2.

**Emergence of type-II Dirac points.** Given the elementary nature of the individual resonant elements, one may be tempted to assume that nothing peculiar could emerge from the ordinary dipole–dipole interactions between the meta-atoms which are mediated by the electromagnetic field. However, by expressing the interaction Hamiltonian (Eq. (4)) in terms of the $\tilde{\beta}_{\mathbf{q}\tau}$ and $\tilde{\beta}_{\mathbf{q}\tau}^\dagger$ operators that diagonalize the matter Hamiltonian,

$$H_{int}^{(1)} = \hbar \sum_{\tau=\pm} \sum_{\mathbf{q}n} i\Lambda_{\mathbf{q}n\tau}\left(\tilde{\beta}_{\mathbf{q}\tau}^\dagger c_{\mathbf{q}n} + \tilde{\beta}_{\mathbf{q}\tau}^\dagger c_{-\mathbf{q}n}^\dagger\right) + \text{H.c.}, \quad (6)$$

we find that complex non-local interactions, which arise from strong multiple-scattering in the bipartite structure, result in a non-trivial winding of the light-matter coupling as a function of wavevector direction

$$\Lambda_{\mathbf{q}n\tau} \propto \xi_{\mathbf{q}n}\left(\phi_n^* e^{i\varphi_{\mathbf{q}}} + \tau\phi_n\right). \quad (7)$$

Naively, one may expect all of the band crossings in Fig. 2b to be avoided as a result of the hybridization between the collective-dipole and photonic modes, as it is a characteristic feature of polaritonic systems[48,49]. Indeed, this is the case for the crossings with the upper quasistatic band where $\Lambda_{\mathbf{q}0+} \propto (e^{i\varphi_{\mathbf{q}}} + 1)$ (see red line in Fig. 3a) due to the constructive interference between the sublattices of this bright (↑↑) configuration (see Fig. 3b, c). This results in a large anticrossing for all wavevector directions, as observed in Fig. 2c. In stark contrast, for the lower quasistatic band the coupling constant is significantly reduced $\Lambda_{\mathbf{q}0-} \propto (e^{i\varphi_{\mathbf{q}}} - 1)$ (see blue line in Fig. 3a) due to the destructive interference between the sublattices of this dark (↑↓) configuration (see Fig. 3e). Consequently, this results in a small anticrossing for a general wavevector direction.

Crucially, however, the light–matter interaction for the lower quasistatic band completely vanishes ($\Lambda_{\mathbf{q}0-} = 0$) along the high-symmetry $\Gamma-K(K')$ directions, where $\varphi_{\mathbf{q}} = 0$, due to the complete destructive interference between the two sublattices (see Fig. 3d). As a result, along these high-symmetry directions the crossings are protected, leading to six inequivalent Dirac points emerging in the polariton spectrum—we call these satellite Dirac points (SDPs) to distinguish them from the CDPs. As we will see below, these SDPs belong to the type-II class of 2D Dirac points where the dispersion takes the form of a critically tilted Dirac cone (see inset of Fig. 2c), characterized by open, hyperbolic isofrequency contours.

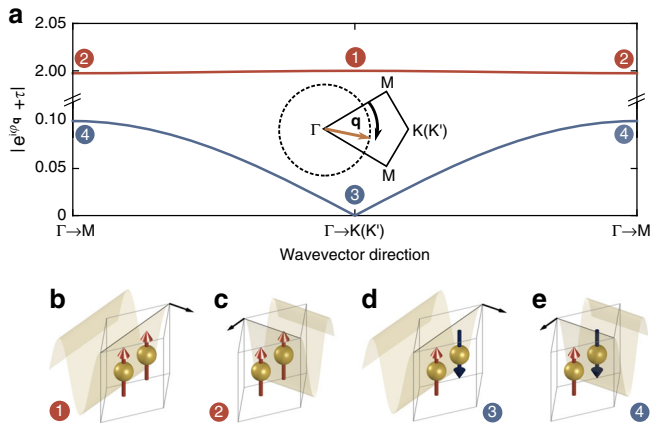

**Fig. 3** Non-trivial winding in the light–matter interaction. **a** Dependence of the magnitude of the light–matter coupling constant $|\Lambda_{\mathbf{q}0\tau}| \propto |e^{i\varphi_{\mathbf{q}}} + \tau|$ on the direction of $\mathbf{q}$ from $\Gamma-M$ to $\Gamma-K(K')$ (see inset), for the upper (red line) and lower (blue line) quasistatic bands. Plots obtained with $|\mathbf{q}| = |\mathbf{K}|/2$. **b**–**e** Schematics of the bright ($\uparrow\uparrow$) and dark ($\uparrow\downarrow$) configurations of the two sublattices interacting with the photonic mode which is indicated by the field profile. Panels **b** and **d** represent the crossings labeled '1' and '3' in Fig. 2 along the $\Gamma-K(K')$ directions, respectively, while **c** and **e** represent the crossings labeled '2' and '4' along the $\Gamma-M$ directions, respectively. Crucially, the light–matter interaction strength for the dark mode vanishes ($|\Lambda_{\mathbf{q}0-}| = 0$) along the $\Gamma-K(K')$ directions due to the complete destructive interference between the two sublattices (see **d**), leading to the emergence of six inequivalent type-II Dirac points in the polariton spectrum

**Effective Hamiltonian in the matter subspace.** To explore the nature of the polaritons in the vicinity of the different Dirac points, we first neglect non-resonant terms in the matter Hamiltonian and perform a unitary Schrieffer–Wolff transformation[50] on $H_{pol}$ to integrate out the photonic degrees of freedom (see Methods for details). Finally, we extract the two-band Hamiltonian in the matter sublattice space

$$\overline{H}_{mat} = H_{mat} - 2\hbar \sum_{\mathbf{q}n} \frac{\xi_{\mathbf{q}n}^2 \omega_{\mathbf{q}n}^{ph}}{\left(\omega_{\mathbf{q}n}^{ph}\right)^2 - \tilde{\omega}_{\mathbf{q}}^2}$$
$$\left(a_{\mathbf{q}}^\dagger a_{\mathbf{q}} + b_{\mathbf{q}}^\dagger b_{\mathbf{q}} + \phi_n^2 b_{\mathbf{q}}^\dagger a_{\mathbf{q}} + \phi_n^{*2} a_{\mathbf{q}}^\dagger b_{\mathbf{q}}\right). \quad (8)$$

Diagonalizing the two-band Hamiltonian (Eq. (8)) leads to an effective dispersion (see Methods) which provides an excellent description of the polaritons as indicated by the orange dashed lines in Fig. 2c–e. Finally, we expand the two-band Hamiltonian (Eq. (8)) up to quadratic in $\mathbf{k}$ and obtain the effective Hamiltonian near the K point $\overline{H}_K^{eff} = \sum_{\mathbf{k}} \psi_{\mathbf{k}}^\dagger \overline{\mathcal{H}}_{K,\mathbf{k}}^{eff} \psi_{\mathbf{k}}$ (see Supplementary Note 2 for derivation) with Bloch Hamiltonian

$$\overline{\mathcal{H}}_{K,\mathbf{k}}^{eff} = \hbar \overline{\omega}_0 \mathbb{1}_2 - \hbar \overline{v} \boldsymbol{\sigma} \cdot \mathbf{k} + \hbar \overline{t} (\boldsymbol{\sigma}^* \cdot \mathbf{k})^{\circ 2} - \hbar D |\mathbf{k}|^2 \mathbb{1}_2. \quad (9)$$

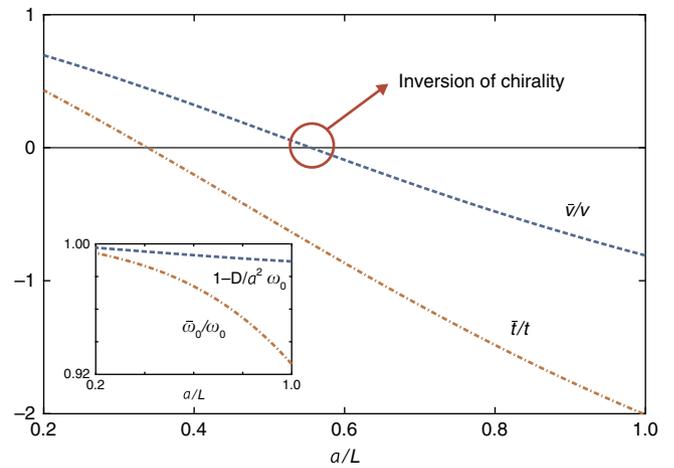

**Fig. 4** Tunable parameters in the effective Hamiltonian. Dependence of the parameters in the effective polariton Hamiltonian (Eq. (9)) on the inverse cavity height. The blue dashed line shows the variation of the group velocity $\overline{v}$ which changes sign at the critical cavity height $L_c$, leading to the inversion of chirality. The orange dot-dashed line shows the variation of the trigonal warping parameter $\overline{t}$ which becomes dominant close to criticality. These parameters have been normalized to $v = 3\Omega a/2$ and $t = 3\Omega a^2/8$ which are the group velocity and trigonal warping parameters, respectively, in the absence of image dipoles and light–matter interactions. The orange dot-dashed line in the inset shows the variation of the CDP frequency $\overline{\omega}_0$, while the blue dashed line in the inset shows the variation of the wavevector-dependent diagonal term $D$. Plots obtained with $\omega_{\mathbf{K}0}^{ph} = 2.5\omega_0$ and $\Omega = 0.01\omega_0$

Similarly, the effective Hamiltonian near the K' point is given by $\overline{\mathcal{H}}_{K',\mathbf{k}}^{eff} = (\overline{\mathcal{H}}_{K,-\mathbf{k}}^{eff})^*$. In Eq. (9), the resonant frequency $\overline{\omega}_0$, group velocity $\overline{v}$, and trigonal warping parameter $\overline{t}$, now encode non-trivial contributions from the hybridization with the photonic environment. There is also an additional wavevector-dependent diagonal term parametrized by $D$, which breaks the symmetry between the upper and lower polariton bands. The dependence of these parameters on the cavity height is shown in Fig. 4 (see Methods for analytical expressions). To leading order in $\mathbf{k}$, one can observe that the effective Hamiltonian (Eq. (9)) near the CDP is equivalent to a 2D massless Dirac Hamiltonian. Therefore, the polariton CDPs remain in the type-I class and are robust against the coupling with the photonic environment—this is not surprising given that their physical origin is intrinsically linked to the lattice structure alone, which is preserved here.

To elucidate the nature of the SDPs, we expand the effective Hamiltonian (Eq. (9)) near one of the SDPs located at $\mathbf{K}_S = (\overline{v}/\overline{t}, 0)$ and obtain

$$\overline{\mathcal{H}}_{K_S,\mathbf{k}'}^{eff} = \hbar \left(\overline{\omega}_0 - \frac{D\overline{v}^2}{\overline{t}^2} - \frac{2D\overline{v}}{\overline{t}} k_x'\right) \mathbb{1}_2 + \hbar \boldsymbol{\sigma}^* \cdot \overline{\mathbf{v}} \cdot \mathbf{k}', \quad (10)$$

where $\mathbf{k}'$ measures the deviation from $\mathbf{K}_S$ and $\overline{\mathbf{v}} = \overline{v}\begin{pmatrix} 1 & 0 \\ 0 & 3 \end{pmatrix}$ is the velocity tensor. Apart from a global energy shift, the effective Hamiltonian (Eq. (10)) near the SDP takes the form of a generalized 2D massless Dirac Hamiltonian $\mathcal{H}_\mathbf{k} = \sum_{i=x,y} \hbar u_i k_i \mathbb{1}_2 + \sum_{i=x,y} \hbar v_i k_i \sigma_i$. If the parameters $u_i$ and $v_i$ satisfy the condition $u_x^2/v_x^2 + u_y^2/v_y^2 < 1$, then the Dirac cone becomes tilted and anisotropic[51] but still belongs to the type-I class with closed isofrequency contours. However, the condition $u_x^2/v_x^2 + u_y^2/v_y^2 > 1$ defines a distinct type-II class of 2D Dirac points, giving rise to a critically tilted Dirac cone with open, hyperbolic isofrequency contours. Hence, the type-I and type-II classes are related via a Lifshitz transition in the topology of the isofrequency contours. Indeed, since we have $u_y = 0$ and $u_x^2/v_x^2 = 4D^2/\overline{t}^2 > 1$, the SDPs belong to the type-II class of 2D Dirac points. Furthermore, since the Hamiltonian (Eq. (10)) is expressed in terms of $\vec{\sigma}^*$, the pseudospin winds in the opposite direction around the SDPs as compared to the CDP, and therefore the type-II SDPs located along the $\Gamma-$K directions are sources of $-\pi$ Berry flux. As required by time-reversal symmetry, the SDPs located along the $\Gamma-$K' directions are sources of $\pi$ Berry flux (opposite to the CDP located at the K' point).

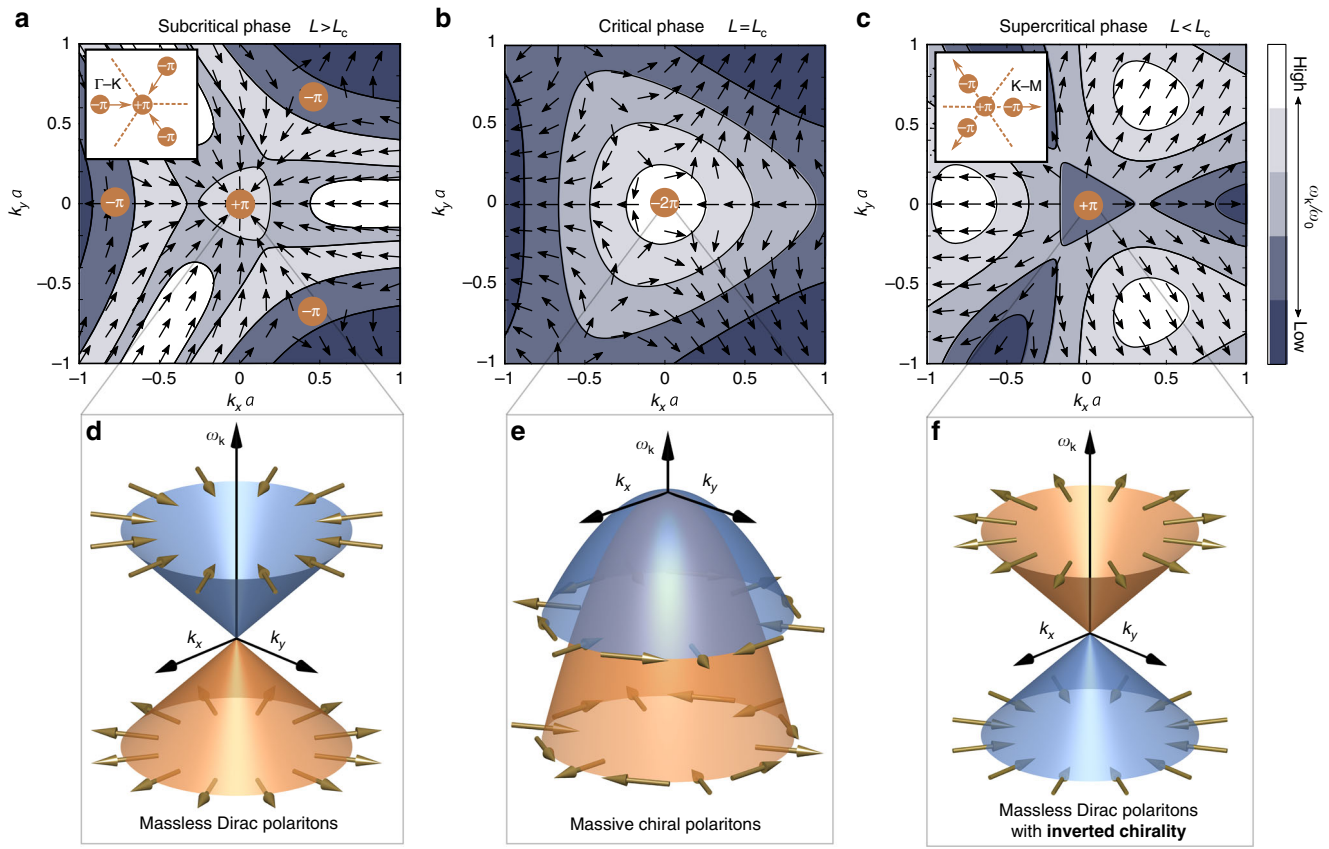

**Fig. 5** Merging of type-I and type-II Dirac points with chirality inversion. **a–c** Pseudospin vector field and isofrequency contours (see Methods for the specific isofrequency values) for the upper polariton band near the K point for **a** subcritical ($L = 2.3a$), **b** critical ($L = L_c = 1.78a$), and **c** supercritical ($L = 1.4a$) cavity heights, respectively, as obtained from $\overline{H}_{\mathrm{mat}}$. The Dirac points (corresponding to vortices) are depicted by orange circles along with their associated Berry flux. Before criticality, three type-II SDPs ($-\pi$ Berry flux) are driven towards the type-I CDP ($\pi$ Berry flux) along the $\Gamma-$K directions as the cavity height is reduced (see inset in **a**). At the critical cavity height $L_c$, they merge together forming a quadratic band-degeneracy with combined Berry flux of $-2\pi$ (see **b**). After criticality, the type-II SDPs re-emerge and are driven past the type-I CDP along the K–M directions (see inset in **c**). After a small decrease in cavity height, these SDPs annihilate other SDPs that migrate along the opposite direction and have opposite Berry flux, leaving only the type-I CDP remaining in the spectrum with $\pi$ Berry flux (see **c**). **d-f**, Effective polariton spectrum near the K point to leading order in **k**. The colors of the bands correspond to the chirality of the Dirac polaritons as defined in the main text, where the orange and blue bands indicate a chirality of $+1$ and $-1$, respectively. The spinor eigenstates, represented by pseudospin vectors (gold arrows), describe **d** massless Dirac polaritons with a linear dispersion and Berry phase of $\pi$, **e** massive chiral polaritons with a parabolic dispersion and Berry phase of $-2\pi$, and **f** massless Dirac polaritons with a linear dispersion and Berry phase of $\pi$, but with inverted chirality. All pseudospin and contour plots are obtained with $\omega_{\mathrm{K0}}^{\mathrm{ph}} = 2.5\omega_0$ and $\Omega = 0.01\omega_0$.

**Manipulation of type-I and type-II Dirac points**. We have thus demonstrated that the honeycomb metasurface simultaneously exhibits two distinct species of massless Dirac polaritons, namely type-I and type-II. In contrast to the type-I CDPs, the existence of the type-II SDPs is intrinsically linked to the hybridization between the light and matter degrees of freedom, and thus one can manipulate their location within the Brillouin zone by simply modifying the light–matter interaction via the cavity height. As a result, the polariton spectrum evolves into qualitatively distinct phases as highlighted in Fig. 2c–e. To elucidate the differences between these phases, we study the spinor eigenstates (see Methods) of the two-band Hamiltonian (Eq. (8)). In Fig. 5a–c we plot the pseudospin vector field near the K point for different cavity heights, along with schematically depict the location of the Dirac points, along with their associated Berry flux. Finally, in Fig. 5d–f, we illustrate the corresponding effective polariton spectrum to leading order in **k**. Note that similar analysis can be performed at the K' point.

In the subcritical phase ($L > L_c$), three type-II SDPs are located along the $\Gamma-$K directions, each with $-\pi$ Berry flux surrounding a type-I CDP with $\pi$ Berry flux (see Fig. 5a). To leading order in **k**, the polariton spectrum disperses linearly about the type-I CDPs (see Fig. 2c) forming an isotropic Dirac cone with a group velocity $\overline{v}$ that is tunable via the cavity height (see Fig. 4). Here, the effective Hamiltonian (Eq. (9)) is equivalent to a 2D massless Dirac Hamiltonian with spinor eigenstates $|\psi_{\mathbf{k}\tau}\rangle = (1, -\tau e^{i\theta_{\mathbf{k}}})^{\mathrm{T}}/\sqrt{2}$. These represent massless Dirac polaritons with chirality $\langle \psi_{\mathbf{k}\tau}|\boldsymbol{\sigma} \cdot \hat{\mathbf{k}}|\psi_{\mathbf{k}\tau}\rangle = -\tau$, resulting in a pseudospin that winds once around the CDP and a topological Berry phase of $\pi$ (see Fig. 5d).

At the critical cavity height ($L = L_c$), the group velocity of the massless Dirac polaritons vanishes $\overline{v}(L_c) = 0$ (see Fig. 4) as the type-II SDPs merge with the type-I CDP, forming a quadratic band-degeneracy (see Fig. 2d) with combined $-2\pi$ Berry flux (see

Fig. 5b). The leading order term in the effective Hamiltonian (Eq. (9)) is now quadratic in $\mathbf{k}$ with corresponding spinor eigenstates $|\psi_{\mathbf{k}\tau}\rangle = (1, -\tau e^{-i2\theta_{\mathbf{k}}})^T/\sqrt{2}$. Therefore, during this critical merging transition, the massless Dirac polaritons morph into massive chiral polaritons with qualitatively distinct physical properties. These include a parabolic spectrum and chirality $\langle\psi_{\mathbf{k}\tau}|(\sigma^* \cdot \hat{\mathbf{k}})^{\circ 2}|\psi_{\mathbf{k}\tau}\rangle = -\tau$, resulting in a pseudospin that winds twice as fast compared to the subcritical phase and a topological Berry phase of $-2\pi$ (see Fig. 5e).

Since the point-group symmetry is preserved, the type-II SDPs do not annihilate the type-I CDP, but they re-emerge and continue to migrate along the K−M directions as the cavity height is reduced past criticality ($L < L_c$) (see inset of Fig. 5c). After a small decrease in cavity height, these SDPs annihilate with other SDPs that migrate along the opposite direction and have opposite Berry flux. This topological transition leaves only the type-I CDP remaining in the polariton spectrum with $\pi$ Berry flux (see Figs. 2e and 5c).

In this supercritical phase, we recover the linear dispersion near the type-I CDP to leading order in $\mathbf{k}$ (see Fig. 2e), and the effective Hamiltonian (Eq. (9)) is equivalent to a 2D massless Dirac Hamiltonian with corresponding spinor eigenstates $|\psi_{\mathbf{k}\tau}\rangle = (1, \tau e^{i\theta_{\mathbf{k}}})^T/\sqrt{2}$. Remarkably, massless Dirac polaritons thus re-emerge past criticality with an environment-induced inversion of chirality $\langle\psi_{\mathbf{k}\tau}|\sigma \cdot \hat{\mathbf{k}}|\psi_{\mathbf{k}\tau}\rangle = \tau$ (see Fig. 5f). Physically, this corresponds to a $\pi$-rotation in the relative phase between the dipole oscillations on the two inequivalent sublattices, which is also accompanied by a $\pi$-rotation in the isofrequency domains (compare Fig. 5a and Fig. 5c).

We emphasize that it is the chirality of massless Dirac fermions that is responsible for most of the remarkable properties of monolayer graphene, including the Klein tunneling phenomenon[13]. Consequently, this unique environment-induced inversion of chirality could give rise to unconventional phenomena such as anomalous Klein transport. For example, near the K point, the right-propagating polaritons correspond to an antisymmetric dipole configuration $|\psi_{\mathbf{k}\tau}\rangle = (1, -1)^T/\sqrt{2}$ in the subcritical phase and to a symmetric configuration $|\psi_{\mathbf{k}\tau}\rangle = (1, 1)^T/\sqrt{2}$ in the supercritical phase. Thus, due to the orthogonality between these two spinor eigenstates, the inversion of chirality removes the channel responsible for the perfect transmission in the conventional Klein tunneling effect[13] (see Fig. 5d, f). Such a scenario could be realized in a simple setup characterized by two neighboring regions with different cavity heights.

As a side remark, we note that the polariton spectrum near criticality bears some resemblance with the low-energy spectrum of bilayer graphene with its central Dirac point and three satellite Dirac points, which all belong to the type-I class[52,53]. However, given the type-II nature of the polariton SDPs, the topology of the polariton isofrequency contours are markedly different from that of the bilayer spectrum. This is further highlighted at criticality where the polariton bands have the same curvature, which is in stark contrast to the electronic bands in bilayer graphene.

We also note that recent works explored the possibility to manipulate the (3+1) type-I Dirac points in bilayer graphene through the application of lattice deformations[54–57], leading to the merging and annihilation of pairs of Dirac points. In addition, a multimerging transition of all (3+1) type-I Dirac points has been proposed theoretically within tight-binding models involving the artificial tuning of third-nearest-neighbor hopping amplitudes in a graphene-like honeycomb structure[58–61]. However, these proposals have no physical realization so far. In stark contrast, the imminently realizable metasurfaces in our work enable the exploration of rich Dirac phases with ease by simply modifying the photonic environment via an enclosing cavity.

As a final remark, we briefly comment on how one might probe the Dirac physics presented in this work. Given that the Dirac points exist in a polaritonic excitation spectrum, one must drive the system with photons at the required frequency in order to probe them. In fact, both of the type-I and type-II Dirac points lie outside of the light-line and therefore one must overcome the momentum mismatch with photons. The specific experimental technique that one would employ will depend on the nature of the metasurface and the corresponding frequency regime. For example, techniques for plasmonic systems have traditionally involved coupling via evanescent waves with prisms, gratings, and local scatterers[62], or more recent techniques such as non-linear wave-mixing[63]. In contrast, realizations in the microwave regime can be probed using point-like antenna sources and detectors[33]. In fact, microwave metamaterials are proving to be a versatile platform for exploring Dirac/Weyl physics, as one can directly probe the field distributions using near-field scanning techniques[33], and thus one could directly probe the environment-induced chirality inversion predicted here.

## Discussion

To conclude, we have revealed rich and unique Dirac physics that emerges even in the most elementary honeycomb metasurfaces. In particular, we have unveiled the simultaneous existence of both type-I and type-II massless Dirac polaritons, where the latter emerge from a non-trivial winding in the light–matter interaction. We would like to emphasize that it is this unique physical origin of the type-II SDPs, together with the truly 2D nature of the metasurface, that enables one to qualitatively modify the fundamental properties of these emergent Dirac polaritons by manipulating the surrounding photonic environment alone. This stands in stark contrast to conventional artificial graphene systems where the fundamental properties are dictated by the lattice structure. Therefore, exploiting the rich tunability of the polariton spectrum with the environment offers a new paradigm that opens a variety of opportunities to explore unique Dirac-related physics at the subwavelength scale.

For example, one can simultaneously probe the dynamics of type-I and type-II massless Dirac quasiparticles, where the latter are predicted to exhibit intriguing anomalous refraction behavior[34,35]. Furthermore, the environment-induced redshift of the CDP frequency $\overline{\omega}_0$ (see Fig. 4) will allow the investigation of polaritonic Klein tunneling through interfaces separating regions with different cavity heights. Consequently, negative refraction can be induced by simple variations in the cavity height, which could be exploited in novel schemes for guiding and manipulating light at the subwavelength scale, including polaritonic Veselago lensing[64,65]. Moreover, the tunable group velocity will enable the exploration of velocity barriers for the unprecedented guiding and localization of massless Dirac quasiparticles[66,67], which is extremely difficult to achieve in real graphene. One could also combine the effects of the environment with inhomogeneous strain deformations, giving rise to unique pseudomagnetic-related effects, including the intriguing ability to induce a pseudo-Landau level spectrum for polaritons that can be qualitatively tuned via the cavity height. Finally, the ability to controllably invert the chirality of the massless Dirac polaritons opens new perspectives for anomalous pseudorelativistic transport through interfaces separating regions in distinct polaritonic phases.

## Methods

**Derivation of the polaritonic Hamiltonian.** The cavity-embedded metasurface is composed of a honeycomb array of identical meta-atoms located at $\mathbf{R}_A = \mathbf{R} + a\hat{\mathbf{y}} + \frac{L}{2}\hat{\mathbf{z}}$ and $\mathbf{R}_B = \mathbf{R} - a\hat{\mathbf{y}} + \frac{L}{2}\hat{\mathbf{z}}$ on the A and B sublattices, respectively.

Here, $\mathbf{R} = l_1\mathbf{a}_1 + l_2\mathbf{a}_2$ is an in-plane lattice translation vector with primitive vectors $\mathbf{a}_1$ and $\mathbf{a}_2$ (see Fig. 1) and integers $l_1$ and $l_2$. Each meta-atom is modeled by a single dynamical degree of freedom $h$ (with dimensions of length), where the electric-dipole moment associated with its fundamental eigenmode is $\mathbf{p} = -Qh\hat{z}$, with effective charge $Q$. The Coulomb potential energy between two dipole moments $\mathbf{p}$ and $\mathbf{p}'$ at generic positions $\mathbf{r}$ and $\mathbf{r}'$, respectively, is given by

$$V_{\text{Coul}} = \frac{\mathbf{p} \cdot \mathbf{p}' - 3(\mathbf{p} \cdot \hat{\mathbf{n}})(\mathbf{p}' \cdot \hat{\mathbf{n}})}{4\pi\varepsilon_0 |\mathbf{r} - \mathbf{r}'|^3}, \tag{11}$$

where $\hat{\mathbf{n}} = (\mathbf{r} - \mathbf{r}')/|\mathbf{r} - \mathbf{r}'|$ and $\varepsilon_0$ is the vacuum permittivity.

The presence of the perfectly conducting metallic plates, placed at $z = 0$ and $z = L$, modifies the boundary conditions on the scalar potential and, therefore, the Coulomb interaction between the meta-atoms. Using the method of images to ensure the vanishing of the scalar potential at the cavity walls[68], we introduce an infinite series of image dipoles located outside the cavity at positions $\mathbf{R}_s + lL\hat{z}$, where $s = $ A, B labels the two sublattices and $l$ is a non-zero integer. Noting that the Coulomb potential energy between a real and image dipole is 1/2 of that given by Eq. (11)[69], the matter Hamiltonian within the nearest-neighbor approximation reads

$$
\begin{aligned}
H_{\text{mat}} = & \sum_{s=\text{A,B}} \sum_{\mathbf{R}_s} \left( \frac{\Pi_{\mathbf{R}_s}^2}{2M} + \frac{M}{2}\omega_0^2 h_{\mathbf{R}_s}^2 \right) \\
& + \frac{Q^2}{4\pi\varepsilon_0 a^3} \sum_{\mathbf{R}_B} \sum_{j=1}^{3} h_{\mathbf{R}_B} h_{\mathbf{R}_B + \mathbf{e}_j} \\
& - \frac{Q^2}{8\pi\varepsilon_0 a^3} \sum_{s=\text{A,B}} \sum_{\mathbf{R}_s} \sum_{l=-\infty}^{+\infty} {}^{\prime} 2 \left|\frac{a}{lL}\right|^3 h_{\mathbf{R}_s}^2 \\
& - \frac{Q^2}{8\pi\varepsilon_0 a^3} \sum_{\mathbf{R}_B} \sum_{j=1}^{3} \sum_{l=-\infty}^{+\infty} {}^{\prime} \frac{2\left|\frac{a}{lL}\right|^2 - 1}{\left(1 + \left|\frac{a}{lL}\right|^2\right)^{\frac{5}{2}}} h_{\mathbf{R}_B} h_{\mathbf{R}_B + \mathbf{e}_j} \\
& - \frac{Q^2}{8\pi\varepsilon_0 a^3} \sum_{\mathbf{R}_A} \sum_{j=1}^{3} \sum_{l=-\infty}^{+\infty} {}^{\prime} \frac{2\left|\frac{a}{lL}\right|^2 - 1}{\left(1 + \left|\frac{a}{lL}\right|^2\right)^{\frac{5}{2}}} h_{\mathbf{R}_A} h_{\mathbf{R}_A - \mathbf{e}_j}
\end{aligned} \tag{12}
$$

where the primed summations exclude the $l = 0$ term. Here, $\Pi_{\mathbf{R}_s}$ is the conjugate momentum to the dynamical coordinate $h_{\mathbf{R}_s}$ corresponding to the meta-atom located at $\mathbf{R}_s$, and $M$ is an effective mass. Next, we introduce the bosonic operators

$$a_{\mathbf{R}_A} = \sqrt{\frac{M\omega_0}{2\hbar}} h_{\mathbf{R}_A} + i\sqrt{\frac{1}{2\hbar M\omega_0}} \Pi_{\mathbf{R}_A} \tag{13}$$

and

$$b_{\mathbf{R}_B} = \sqrt{\frac{M\omega_0}{2\hbar}} h_{\mathbf{R}_B} + i\sqrt{\frac{1}{2\hbar M\omega_0}} \Pi_{\mathbf{R}_B} \tag{14}$$

that annihilate quanta of the fundamental eigenmode on the meta-atom located at $\mathbf{R}_A$ and $\mathbf{R}_B$, respectively, and satisfy the commutation relations $[a_{\mathbf{R}}, a_{\mathbf{R}'}^{\dagger}] = \delta_{\mathbf{R}\mathbf{R}'}$, $[b_{\mathbf{R}}, b_{\mathbf{R}'}^{\dagger}] = \delta_{\mathbf{R}\mathbf{R}'}$, and $[a_{\mathbf{R}}, b_{\mathbf{R}'}^{\dagger}] = 0$. In terms of these operators, the matter Hamiltonian (Eq. (12)) reads

$$
\begin{aligned}
H_{\text{mat}} = & \hbar\omega_0 \sum_{\mathbf{R}_A} a_{\mathbf{R}_A}^{\dagger} a_{\mathbf{R}_A} + \hbar\omega_0 \sum_{\mathbf{R}_B} b_{\mathbf{R}_B}^{\dagger} b_{\mathbf{R}_B} \\
& + \hbar\Omega(1 - \mathcal{I}) \sum_{\mathbf{R}_B} \sum_{j=1}^{3} \left[ b_{\mathbf{R}_B}^{\dagger} \left( a_{\mathbf{R}_B + \mathbf{e}_j} + a_{\mathbf{R}_B + \mathbf{e}_j}^{\dagger} \right) + \text{H.c.} \right] \\
& - \frac{\hbar}{2}\omega_0 \mathcal{S} \sum_{\mathbf{R}_A} \left[ a_{\mathbf{R}_A}^{\dagger} \left( a_{\mathbf{R}_A} + a_{\mathbf{R}_A}^{\dagger} \right) + \text{H.c.} \right] \\
& - \frac{\hbar}{2}\omega_0 \mathcal{S} \sum_{\mathbf{R}_B} \left[ b_{\mathbf{R}_B}^{\dagger} \left( b_{\mathbf{R}_B} + b_{\mathbf{R}_B}^{\dagger} \right) + \text{H.c.} \right]
\end{aligned} \tag{15}
$$

where $\Omega = Q^2/8\pi\varepsilon_0 M\omega_0 a^3$ parametrizes the strength of the nearest-neighbor Coulomb interaction, and the parameters

$$\mathcal{S} = 4\sum_{l=1}^{\infty} \left(\frac{a}{lL}\right)^3, \quad \mathcal{I} = 2\sum_{l=1}^{\infty} \frac{2\left(\frac{lL}{a}\right)^2 - 1}{\left[1 + \left(\frac{lL}{a}\right)^2\right]^{\frac{5}{2}}} \tag{16}$$

encode renormalizations due to the cavity-induced image dipoles. We apply Born–von Kármán boundary conditions over a lattice with $\mathcal{N} \gg 1$ unit cells and introduce the Fourier transform of the bosonic operators $a_{\mathbf{R}_A} = \mathcal{N}^{-1/2} \sum_{\mathbf{q}} a_{\mathbf{q}} e^{i\mathbf{q}\cdot\mathbf{R}_A}$ and $b_{\mathbf{R}_B} = \mathcal{N}^{-1/2} \sum_{\mathbf{q}} b_{\mathbf{q}} e^{i\mathbf{q}\cdot\mathbf{R}_B}$, which transforms the matter Hamiltonian (Eq. (15)) into

the local and block-diagonal form

$$
\begin{aligned}
H_{\text{mat}} = & \hbar \sum_{\mathbf{q}} (\omega_0 - \Omega\mathcal{S}) \left( a_{\mathbf{q}}^{\dagger} a_{\mathbf{q}} + b_{\mathbf{q}}^{\dagger} b_{\mathbf{q}} \right) \\
& + \hbar \sum_{\mathbf{q}} \Omega(1 - \mathcal{I}) \left[ f_{\mathbf{q}} b_{\mathbf{q}}^{\dagger} \left( a_{\mathbf{q}} + a_{-\mathbf{q}}^{\dagger} \right) + \text{H.c.} \right] \\
& - \hbar \sum_{\mathbf{q}} \frac{1}{2}\Omega\mathcal{S} \left( a_{\mathbf{q}}^{\dagger} a_{-\mathbf{q}}^{\dagger} + b_{\mathbf{q}}^{\dagger} b_{-\mathbf{q}}^{\dagger} + \text{H.c.} \right).
\end{aligned} \tag{17}
$$

In the main text, we do not present the non-resonant terms (e.g., $b_{\mathbf{q}}^{\dagger} a_{-\mathbf{q}}^{\dagger}$), leading to Eq. (1) where $\tilde{\omega}_0 = \omega_0 - \Omega\mathcal{S}$ and $\tilde{\Omega} = \Omega(1 - \mathcal{I})$.

In the Coulomb gauge, the light–matter interaction is described by the minimal-coupling Hamiltonian[46] which, within the dipole approximation, reads

$$H_{\text{int}} = \underbrace{\frac{Q}{M} \sum_{s=\text{A,B}} \sum_{\mathbf{R}_s} \Pi_{\mathbf{R}_s} A_z(\mathbf{R}_s)}_{H_{\text{int}}^{(1)}} + \underbrace{\frac{Q^2}{2M} \sum_{s=\text{A,B}} \sum_{\mathbf{R}_s} A_z^2(\mathbf{R}_s)}_{H_{\text{int}}^{(2)}}, \tag{18}$$

where we have used $\Pi_{\mathbf{R}_s} = \Pi_{\mathbf{R}_s}\hat{z}$. The vector potential can be decomposed into transverse electric (TE) and transverse magnetic (TM) modes of the cavity. However, the photons corresponding to the TE modes have an in-plane polarization, and therefore only TM modes contribute to the $z$-component of the vector potential

$$
\begin{aligned}
A_z(\mathbf{r}, z) = & \sum_{\mathbf{q}mn} \sqrt{\frac{\hbar}{\varepsilon_0 N_m \mathcal{N} A L \omega_{\mathbf{q}mn}^{\text{ph}}}} \frac{|\mathbf{q} - \mathbf{G}_n|}{|\mathbf{q} - \mathbf{G}_n + \frac{m\pi}{L}\hat{z}|} \cos\left(\frac{m\pi}{L} z\right) \\
& \left[ c_{\mathbf{q}mn} e^{i(\mathbf{q} - \mathbf{G}_n)\cdot\mathbf{r}} + c_{\mathbf{q}mn}^{\dagger} e^{-i(\mathbf{q} - \mathbf{G}_n)\cdot\mathbf{r}} \right],
\end{aligned} \tag{19}
$$

where $\mathcal{A} = 3\sqrt{3}a^2/2$ is the area of a unit cell and $N_m = 1 + \delta_{m0}$. The bosonic operator $c_{\mathbf{q}mn}^{\dagger}$ creates a TM photon with wavevector $\mathbf{q}$ in the first Brillouin zone and dispersion $\omega_{\mathbf{q}mn}^{\text{ph}} = c|\mathbf{q} - \mathbf{G}_n + \hat{z}m\pi/L|$. Here, $\mathbf{G}_n = n_1\mathbf{b}_1 + n_2\mathbf{b}_2$ is a reciprocal lattice vector with primitive vectors $\mathbf{b}_1$ and $\mathbf{b}_2$, where $n$ indexes the set of ordered pairs of integers $(n_1, n_2)$, and $m$ is a non-negative integer indexing the different TM cavity modes. Only TM photons with even $m$ couple to the dipoles due to the parity selection rule at the center of the cavity.

Substituting the vector potential (Eq. (19)) into Eq. (18) we obtain the light–matter interaction Hamiltonian given by

$$
\begin{aligned}
H_{\text{int}}^{(1)} = & \hbar \sum_{\mathbf{q}mn} i\xi_{\mathbf{q}mn} \phi_n^* \left( a_{\mathbf{q}}^{\dagger} c_{\mathbf{q}mn} + a_{\mathbf{q}}^{\dagger} c_{-\mathbf{q}mn}^{\dagger} \right) \\
& + \hbar \sum_{\mathbf{q}mn} i\xi_{\mathbf{q}mn} \phi_n \left( b_{\mathbf{q}}^{\dagger} c_{\mathbf{q}mn} + b_{\mathbf{q}}^{\dagger} c_{-\mathbf{q}mn}^{\dagger} \right) + \text{H.c.}
\end{aligned} \tag{20}
$$

and

$$
\begin{aligned}
H_{\text{int}}^{(2)} = & \hbar \sum_{\mathbf{q}mm'nn'} \frac{2\xi_{\mathbf{q}mn}\xi_{\mathbf{q}m'n'}}{\omega_0} \text{Re}\left(\phi_n \phi_{n'}^*\right) \\
& \left( c_{\mathbf{q}mn}^{\dagger} c_{\mathbf{q}m'n'} + c_{\mathbf{q}mn}^{\dagger} c_{-\mathbf{q}m'n'}^{\dagger} \right) + \text{H.c.}
\end{aligned} \tag{21}
$$

The strength of the light–matter interaction is parametrized by

$$\xi_{\mathbf{q}mn} = \omega_0 \mathcal{F}(\omega_{\mathbf{q}mn}^{\text{ph}}) \frac{\omega_{\mathbf{G}0n}^{\text{ph}}}{\omega_{\mathbf{q}mn}^{\text{ph}}} \left( \frac{8\pi}{3\sqrt{3}N_m} \frac{a}{L} \frac{\Omega}{\omega_{\mathbf{q}mn}^{\text{ph}}} \right)^{\frac{1}{2}}, \tag{22}$$

where, to take into account the finite size of the meta-atoms, we have introduced a phenomenological function $\mathcal{F}(\omega_{\mathbf{q}mn}^{\text{ph}})$ that provides a smooth cut-off for the interaction with short-wavelength photonic modes where the dipole approximation breaks down. We choose the phenomenological cut-off function to be of the Fermi–Dirac distribution form

$$\mathcal{F}(\omega_{\mathbf{q}mn}^{\text{ph}}) = \frac{1}{1 + e^{2(\omega_{\mathbf{q}mn}^{\text{ph}} - 3\omega_0)/\omega_0}}, \tag{23}$$

which is smooth enough to avoid spurious artifacts appearing in the polariton spectrum. Finally, the free photonic Hamiltonian of the cavity reads

$$H_{\text{ph}} = \hbar \sum_{\mathbf{q}mn} \omega_{\mathbf{q}mn}^{\text{ph}} c_{\mathbf{q}mn}^{\dagger} c_{\mathbf{q}mn}. \tag{24}$$

In Eqs. (3), (4), (5) in the main text, we only present the contribution from the TEM mode ($m = 0$), dropping the corresponding index. In Supplementary Note 1, we discuss the effect of the higher order ($m \neq 0$) TM cavity modes for larger cavities.

**Hopfield–Bogoliubov diagonalization.** The polariton Hamiltonian $H_{\text{pol}} = H_{\text{mat}} + H_{\text{ph}} + H_{\text{int}}$, where $H_{\text{mat}}$ is given by Eq. (17), $H_{\text{ph}}$ by Eq. (24), and $H_{\text{int}}$ by Eqs. (20) and (21), can be recast into matrix form as $H_{\text{pol}} = \frac{1}{2} \sum_{\mathbf{q}} \Psi_{\mathbf{q}}^{\dagger} \mathcal{H}_{\mathbf{q}}^{\text{pol}} \Psi_{\mathbf{q}}$ where $\Psi_{\mathbf{q}}^{\dagger} = (\psi_{\mathbf{q}}^{\dagger}, C_{\mathbf{q}}^{\dagger}, \psi_{-\mathbf{q}}^{T}, C_{-\mathbf{q}}^{T})$. Here, $\psi_{\mathbf{q}}^{\dagger} = (a_{\mathbf{q}}^{\dagger}, b_{\mathbf{q}}^{\dagger})$ is the spinor creation operator in the matter sublattice space and $C_{\mathbf{q}}^{\dagger} = (c_{\mathbf{q}1}^{\dagger}, c_{\mathbf{q}2}^{\dagger}, \ldots, c_{\mathbf{q}p}^{\dagger}, \ldots, c_{\mathbf{q}N}^{\dagger})$ is the vector of TM photon creation operators, where $p$ indexes the set of ordered triplets of integers $(n_1, n_2, m)$, and $N$ is the total number of photonic operators considered. The Hermitian $[2(N+2)] \times [2(N+2)]$ matrix $\mathscr{H}_{\mathbf{q}}^{\text{pol}}$ can be written in block form as

$$\mathscr{H}_{\mathbf{q}}^{\text{pol}} = \begin{pmatrix} \mathcal{H}_{\mathbf{q}}^{+} & \mathcal{H}_{\mathbf{q}}^{-} - \mathcal{W}_{\mathbf{q}} \\ \left(\mathcal{H}_{\mathbf{q}}^{-} - \mathcal{W}_{\mathbf{q}}\right)^{\dagger} & \left(\mathcal{H}_{-\mathbf{q}}^{+}\right)^{*} \end{pmatrix}, \tag{25}$$

where

$$\mathcal{W}_{\mathbf{q}} = \hbar \text{Diag}\left(\omega_0, \omega_0, \omega_{\mathbf{q}1}^{\text{ph}}, \omega_{\mathbf{q}2}^{\text{ph}}, \ldots, \omega_{\mathbf{q}p}^{\text{ph}}, \ldots, \omega_{\mathbf{q}N}^{\text{ph}}\right) \tag{26}$$

is the $(N+2) \times (N+2)$ diagonal matrix of resonant frequencies of the free oscillators. The $(N+2) \times (N+2)$ block matrices $\mathcal{H}_{\mathbf{q}}^{\pm}$ can be sub-divided into block matrices

$$\mathcal{H}_{\mathbf{q}}^{\pm} = \begin{pmatrix} \mathcal{H}_{\mathbf{q}}^{\text{mat}} & \mathcal{H}_{\mathbf{q}}^{\text{int}} \\ \pm\left(\mathcal{H}_{\mathbf{q}}^{\text{int}}\right)^{\dagger} & \mathcal{H}_{\mathbf{q}}^{\text{ph}} \end{pmatrix}, \tag{27}$$

where the upper-diagonal block

$$\mathcal{H}_{\mathbf{q}}^{\text{mat}} = \hbar \begin{pmatrix} \bar{\omega}_0 & \bar{\Omega}f_{\mathbf{q}}^{*} \\ \bar{\Omega}f_{\mathbf{q}} & \bar{\omega}_0 \end{pmatrix} \tag{28}$$

is the $2 \times 2$ matrix in the matter subspace, and the lower-diagonal block $\mathcal{H}_{\mathbf{q}}^{\text{ph}}$ is the $N \times N$ matrix in the photonic subspace with components

$$\left(\mathcal{H}_{\mathbf{q}}^{\text{ph}}\right)_{pp'} = \hbar\omega_{\mathbf{q}p}^{\text{ph}}\delta_{pp'} + 4\hbar\frac{\xi_{\mathbf{q}p}\xi_{\mathbf{q}p'}}{\omega_0}\text{Re}\left\{\phi_p\phi_{p'}^{*}\right\}. \tag{29}$$

Finally, the off-diagonal block $\mathcal{H}_{\mathbf{q}}^{\text{int}}$ in Eq. (27) is the $2 \times N$ interaction matrix, where the $p$th column reads

$$\left(\mathcal{H}_{\mathbf{q}}^{\text{int}}\right)_{p} = \hbar \begin{pmatrix} i\xi_{\mathbf{q}p}\phi_p \\ i\xi_{\mathbf{q}p}\phi_p^{*} \end{pmatrix}. \tag{30}$$

The polariton Hamiltonian $H_{\text{pol}}$ is diagonalized by a generalized Hopfield–Bogoliubov transformation[48] $T_{\mathbf{q}}$ via $\Psi_{\mathbf{q}} = T_{\mathbf{q}}\chi_{\mathbf{q}}$, where $\chi_{\mathbf{q}}^{\dagger} = (\chi_{\mathbf{q}}^{\dagger}, \chi_{-\mathbf{q}}^{T})$ and $\chi_{\mathbf{q}}^{\dagger} = (\gamma_{\mathbf{q}1}^{\dagger}, \gamma_{\mathbf{q}2}^{\dagger}, \ldots, \gamma_{\mathbf{q}v}^{\dagger}, \ldots, \gamma_{\mathbf{q}N+2}^{\dagger})$. To ensure the invariance of the bosonic commutation relations for the transformed operators, $T_{\mathbf{q}}$ must be a $[2(N+2)] \times [2(N+2)]$ paraunitary matrix[70] that satisfies $T_{\mathbf{q}}\eta_z T_{\mathbf{q}}^{\dagger} = T_{\mathbf{q}}^{\dagger}\eta_z T_{\mathbf{q}} = \eta_z$, where $\eta_z = \sigma_z \otimes \mathbb{1}_2$ and $\sigma_z$ is the Pauli matrix. The transformed bosonic operators $\gamma_{\mathbf{q}v}^{\dagger} = \Psi_{\mathbf{q}}^{\dagger}\eta_z|\Psi_{\mathbf{q}v}\rangle$ and $\gamma_{\mathbf{q}v} = \langle\Psi_{\mathbf{q}v}|\eta_z\Psi_{\mathbf{q}}$ that diagonalize the polariton Hamiltonian as

$$H_{\text{pol}} = \hbar\sum_{\mathbf{q}v}\omega_{\mathbf{q}v}^{\text{pol}}\gamma_{\mathbf{q}v}^{\dagger}\gamma_{\mathbf{q}v}, \tag{31}$$

create and annihilate polaritons in the $v$th band, respectively. The polariton dispersion $\omega_{\mathbf{q}v}^{\text{pol}}$ (black solid lines in Fig. 2c–e) and the corresponding linearly independent eigenvectors $|\Psi_{\mathbf{q}v}\rangle$ (first two columns of $T_{\mathbf{q}}$) are determined from the positive eigenvalue solutions to the non-Hermitian eigenvalue equation $\eta_z\mathscr{H}_{\mathbf{q}}^{\text{pol}}|\Psi_{\mathbf{q}v}\rangle = \hbar\omega_{\mathbf{q}v}^{\text{pol}}|\Psi_{\mathbf{q}v}\rangle$.

**Schrieffer–Wolff transformation.** To obtain an effective two-band Hamiltonian in the matter sublattice space, we neglect non-resonant terms in the matter Hamiltonian (since $\Omega/\omega_0 \ll 1$ for practical realizations of the metasurface), but not in the light–matter interaction Hamiltonian since the photons are not resonant with the collective-dipoles near the corners of the Brillouin zone (see Fig. 2b). Next, we perform a unitary transformation

$$\overline{H}_{\text{pol}} = e^{S}H_{\text{pol}}e^{-S} = H_{\text{pol}} + [S, H_{\text{pol}}] + \frac{1}{2}[S[S, H_{\text{pol}}]] + \ldots \tag{32}$$

and impose the Schrieffer–Wolff condition[50]

$$[S, H_{\text{mat}} + H_{\text{ph}}] = -H_{\text{int}}^{(1)} \tag{33}$$

which eliminates the light–matter interaction to first order in $\xi_{\mathbf{q}mn}$. From Eq. (33), the particular form of the anti-Hermitian operator $S$ reads

$$S = -\sum_{\mathbf{q}mn}\frac{i\xi_{\mathbf{q}mn}}{\omega_{\mathbf{q}mn}^{\text{ph}} - \omega_0}\left(\phi_n^{*}a_{\mathbf{q}}^{\dagger} + \phi_n b_{\mathbf{q}}^{\dagger}\right)c_{\mathbf{q}mn}$$
$$+ \sum_{\mathbf{q}mn}\frac{i\xi_{\mathbf{q}mn}}{\omega_{\mathbf{q}mn}^{\text{ph}} + \omega_0}\left(\phi_n^{*}a_{\mathbf{q}}^{\dagger} + \phi_n b_{\mathbf{q}}^{\dagger}\right)c_{-\mathbf{q}mn}^{\dagger} - \text{H.c.}, \tag{34}$$

where we have used the approximation $|\omega_{\mathbf{q}mn}^{\text{ph}} \pm \omega_0| \gg \Omega|f_{\mathbf{q}}|$ that is valid near the K and K' points. Retaining leading-order terms in $\xi_{\mathbf{q}mn}$, the transformed polariton

Hamiltonian (Eq. (32)) reads

$$\overline{H}_{\text{pol}} \simeq \underbrace{H_{\text{mat}} + \frac{1}{2}\left[S, H_{\text{int}}^{(1)}\right]}_{\overline{H}_{\text{mat}}} + \underbrace{H_{\text{ph}} + H_{\text{int}}^{(2)}}_{\overline{H}_{\text{ph}}}, \tag{35}$$

where the matter and photonic subspaces are decoupled to quadratic order in $\xi_{\mathbf{q}mn}$. Calculating the commutator in Eq. (35) and extracting the Hamiltonian within the matter sublattice space, we obtain the two-band Hamiltonian

$$\overline{H}_{\text{mat}} = H_{\text{mat}} - 2\hbar\sum_{\mathbf{q}mn}\frac{\xi_{\mathbf{q}mn}^2\omega_{\mathbf{q}mn}^{\text{ph}}}{(\omega_{\mathbf{q}mn}^{\text{ph}})^2 - \omega_0^2}\left(a_{\mathbf{q}}^{\dagger}a_{\mathbf{q}} + b_{\mathbf{q}}^{\dagger}b_{\mathbf{q}} + \phi_n^2 b_{\mathbf{q}}^{\dagger}a_{\mathbf{q}} + \phi_n^{*2}a_{\mathbf{q}}^{\dagger}b_{\mathbf{q}}\right). \tag{36}$$

In Eq. (8) in the main text, we only present the contribution from the TEM mode ($m = 0$), dropping the corresponding index. We can recast the Hamiltonian (Eq. (36)) into matrix form as $\overline{H}_{\text{mat}} = \sum_{\mathbf{q}}\psi_{\mathbf{q}}^{\dagger}\overline{\mathcal{H}}_{\mathbf{q}}^{\text{mat}}\psi_{\mathbf{q}}$, with Bloch Hamiltonian

$$\overline{\mathcal{H}}_{\mathbf{q}}^{\text{mat}} = \hbar \begin{pmatrix} W_{\mathbf{q}} & F_{\mathbf{q}}^{*} \\ F_{\mathbf{q}} & W_{\mathbf{q}} \end{pmatrix}. \tag{37}$$

Here $W_{\mathbf{q}} = \tilde{\omega}_0 - \Omega\sum_{mn}\Delta_{\mathbf{q}mn}$ and $F_{\mathbf{q}} = \tilde{\Omega}f_{\mathbf{q}} - \Omega\sum_{mn}\Delta_{\mathbf{q}mn}\phi_n^2$ with

$$\Delta_{\mathbf{q}mn} = \frac{16\pi}{3\sqrt{3}N_m}\frac{a}{L}\frac{\omega_0^2}{(\omega_{\mathbf{q}mn}^{\text{ph}})^2 - \tilde{\omega}_0^2}\left(\frac{\omega_{\mathbf{q}0n}^{\text{ph}}}{\omega_{\mathbf{q}mn}^{\text{ph}}}\right)^2 \mathcal{F}^2(\omega_{\mathbf{q}mn}^{\text{ph}}). \tag{38}$$

Diagonalizing $\overline{H}_{\text{mat}}$ leads to the two-band dispersion $\overline{\omega}_{\mathbf{q}\tau}^{\text{mat}} = W_{\mathbf{q}} + \tau|F_{\mathbf{q}}|$, which is indicated by the orange-dashed lines in Fig. 2c–e. The corresponding spinor eigenstates $|\psi_{\mathbf{q}\tau}\rangle = (1, \tau e^{i\overline{\varphi}_{\mathbf{q}}})^T/\sqrt{2}$, where $\overline{\varphi}_{\mathbf{q}} = \arg(F_{\mathbf{q}})$, can be represented by the pseudospin vector $\mathbf{S}_{\mathbf{q}\tau} = \tau(\cos\overline{\varphi}_{\mathbf{q}}, \sin\overline{\varphi}_{\mathbf{q}}, 0)$ from which we obtain the pseudospin vector field plots in Fig. 5a–c.

**Expansion of the effective two-band Hamiltonian.** Near the K point, the function $\Delta_{\mathbf{q}mn}$, given by Eq. (38), expands as

$$\begin{aligned}\Delta_{\mathbf{k}mn} &\simeq \Delta_{\text{K}mn}^{(0)} - a^2\Delta_{\text{K}mn}^{(1)}\left[(\mathbf{K} - \mathbf{G}_n)_x k_x + (\mathbf{K} - \mathbf{G}_n)_y k_y\right] \\ &+ \frac{1}{2}\left[-a^2\Delta_{\text{K}mn}^{(1)} + a^4\Delta_{\text{K}mn}^{(2)}(\mathbf{K} - \mathbf{G}_n)_x^2\right]k_x^2 \\ &+ \frac{1}{2}\left[-a^2\Delta_{\text{K}mn}^{(1)} + a^4\Delta_{\text{K}mn}^{(2)}(\mathbf{K} - \mathbf{G}_n)_y^2\right]k_y^2 \\ &+ a^4\Delta_{\text{K}mn}^{(2)}(\mathbf{K} - \mathbf{G}_n)_x(\mathbf{K} - \mathbf{G}_n)_y k_x k_y\end{aligned} \tag{39}$$

to quadratic order in $\mathbf{k}$, where the real parameters $\Delta_{\text{K}mn}^{(v)}$ ($v = 0, 1, 2$) depend only on the photon frequencies $\omega_{\text{K}mn}^{\text{ph}}$ at the K point. Collecting the contributions from the degenerate photons (see Supplementary Note 2 for details), we obtain the effective Hamiltonian (Eq. (9)), where parameters are given by

$$\frac{\overline{\omega}_0}{\omega_0} = 1 - \frac{\Omega}{\omega_0}\mathcal{S} - \frac{\Omega}{\omega_0}\sum_{mn}\Delta_{\text{K}mn}^{(0)}, \tag{40}$$

$$\frac{\overline{v}}{v} = 1 - \mathcal{I} - \frac{4\pi}{27}\sum_{mn}A_n\Delta_{\text{K}mn}^{(1)}, \tag{41}$$

$$\frac{\overline{t}}{t} = 1 - \mathcal{I} - \frac{8\pi^2}{81}\sum_{mn}B_n\Delta_{\text{K}mn}^{(2)}, \tag{42}$$

and

$$\frac{D}{\omega_0 a^2} = \frac{\Omega}{\omega_0}\sum_{mn}\left(\frac{4\pi^2}{27}C_n\Delta_{\text{K}mn}^{(2)} - \frac{1}{2}\Delta_{\text{K}mn}^{(1)}\right), \tag{43}$$

with

$$\begin{aligned}A_n &= \frac{\sqrt{3}}{2}(2 - 3n_1)\cos\left[\frac{4\pi}{3}(n_1 + n_2)\right] \\ &+ \frac{1}{2}(6n_2 - 3n_1)\sin\left[\frac{4\pi}{3}(n_1 + n_2)\right],\end{aligned} \tag{44}$$

$$\begin{aligned}B_n &= (3n_1^2 - 6n_2^2 - 6n_1 + 6n_1n_2 + 2)\cos\left[\frac{4\pi}{3}(n_1 + n_2)\right] \\ &+ \sqrt{3}(2n_1 - 4n_2 + 6n_1n_2 - 3n_1^2)\sin\left[\frac{4\pi}{3}(n_1 + n_2)\right],\end{aligned} \tag{45}$$

and

$$C_n = 1 + 3n_1(n_1 - 1) + 3n_2(n_2 - n_1).$$ (46)

For brevity, we retain only the dominant $(m = 0)$ TEM contribution for the plots in Fig. 4, where the coefficients $\Delta_{\mathbf{K}0n}^{(u)}$ in Eq. (39) are given by

$$\Delta_{\mathbf{K}0n}^{(0)} = 2\left(\frac{4\pi}{3\sqrt{3}}\right)\left(\frac{a}{L}\right)\left[\frac{\omega_0^2}{(\omega_{\mathbf{K}0n}^{ph})^2 - \tilde{\omega}_0^2}\right]\mathcal{F}^2(\omega_{\mathbf{K}0n}^{ph}),$$ (47)

$$\Delta_{\mathbf{K}0n}^{(1)} = 4\left(\frac{4\pi}{3\sqrt{3}}\right)^{-1}\left(\frac{a}{L}\right)\left(\frac{\omega_{\mathbf{K}0n}^{ph}}{\omega_0}\right)^2\left[\frac{\omega_0^2}{(\omega_{\mathbf{K}0n}^{ph})^2 - \tilde{\omega}_0^2}\right]^2$$
$$\left\{\mathcal{F}^2(\omega_{\mathbf{K}0n}^{ph}) - \frac{\partial[\mathcal{F}^2(\omega_{\mathbf{K}0n}^{ph})]}{\partial\omega_{\mathbf{K}0n}^{ph}}\frac{(\omega_{\mathbf{K}0n}^{ph})^2 - \tilde{\omega}_0^2}{2\omega_{\mathbf{K}0n}^{ph}}\right\},$$ (48)

and

$$\Delta_{\mathbf{K}0n}^{(2)} = 16\left(\frac{4\pi}{3\sqrt{3}}\right)^{-3}\left(\frac{a}{L}\right)\left(\frac{\omega_{\mathbf{K}0n}^{ph}}{\omega_0}\right)^4\left[\frac{\omega_0^2}{(\omega_{\mathbf{K}0n}^{ph})^2 - \tilde{\omega}_0^2}\right]^3$$
$$\left\{\mathcal{F}^2(\omega_{\mathbf{K}0n}^{ph}) - \frac{\partial[\mathcal{F}^2(\omega_{\mathbf{K}0n}^{ph})]}{\partial\omega_{\mathbf{K}0n}^{ph}}\right.$$
$$\frac{[5(\omega_{\mathbf{K}0n}^{ph})^2 - \tilde{\omega}_0^2][(\omega_{\mathbf{K}0n}^{ph})^2 - \tilde{\omega}_0^2]}{8(\omega_{\mathbf{K}0n}^{ph})^3}$$
$$\left. + \frac{\partial^2[\mathcal{F}^2(\omega_{\mathbf{K}0n}^{ph})]}{(\partial\omega_{\mathbf{K}0n}^{ph})^2}\frac{[(\omega_{\mathbf{K}0n}^{ph})^2 - \tilde{\omega}_0^2]^2}{8(\omega_{\mathbf{K}0n}^{ph})^2}\right\}.$$ (49)

**Isofrequency contour plots.** In ascending order, the normalized frequency values corresponding to the isofrequency contours are 0.98300, 0.98800, 0.99022, and 0.99080 for Fig. 5a, 0.98080, 0.98180, 0.98680, and 0.98300 for Fig. 5b, and finally 0.97215, 0.97278, 0.97350, and 0.97395 for Fig. 5c.

**Data availability**. All relevant data are available from the corresponding authors upon reasonable request.

## Acknowledgements

We would like to thank Ian Hooper for useful discussions and help with the numerical simulations. C.-R.M. and T.J.S. acknowledge financial support from the Engineering and Physical Sciences Research Council (EPSRC) of the United Kingdom. In addition, C.-R.M acknowledges the EPSRC Centre for Doctoral Training in Metamaterials (Grant No. EP/L015331/1) and QinetiQ for additional funding. G.W. acknowledges financial support from Agence Nationale de la Recherche (Project ANR-14-CE26-0005 Q-Meta-Mat) and the Centre National de la Recherche Scientifique through the Projet International de Coopération Scientifique program (Contract No. 6384 APAG). W.L.B. acknowledges the financial support from EPSRC (Grant No. EP/K041150/1) and EU ERC project Photmat (ERC-2016-ADG-742222). E.M. acknowledges financial support from the Leverhulme Trust (Research Project Grant RPG-2015-101), and the Royal Society (International Exchange Grant No. IE140367, Newton Mobility Grants NI160073, and Theo Murphy Award TM160190).

## Author contributions

C.-R.M. performed the theoretical calculations and wrote the manuscript with E.M.; T.J.S. contributed to the theoretical calculations; E.M. and G.W. conceived the project; and E.M. and W.L.B. supervised the project. All authors commented on the manuscript.

## Additional information